\documentclass[graybox]{svmult}

\usepackage{mathptmx}       % selects Times Roman as basic font
\usepackage{helvet}         % selects Helvetica as sans-serif font
\usepackage{courier}        % selects Courier as typewriter font
\usepackage{type1cm}        % activate if the above 3 fonts are
\usepackage{makeidx}         % allows index generation
\usepackage{graphicx}        % standard LaTeX graphics tool
\usepackage{multicol}        % used for the two-column index
\usepackage[bottom]{footmisc}% places footnotes at page bottom

\makeindex             % used for the subject index
\def\kcore{$k$-core}

\def\kisubtree{$(k_i-1)$-ary subtree}

\def\er{Erd\H{o}s-R\'enyi}

\def\ie{i.~e.}

\def\k{\mathbf{k}}

\begin{document}

\title*{Singularities in ternary mixtures of $k$-core percolation}
\author{Davide Cellai and James P.~Gleeson}
\institute{Davide Cellai \at MACSI, Department of Mathematics and Statistics, University of Limerick, Ireland, \email{davide.cellai@ul.ie}
\and James P.~Gleeson \at MACSI, Department of Mathematics and Statistics, University of Limerick, Ireland, \email{james.gleeson@ul.ie}}
\maketitle

\abstract{Heterogeneous $k$-core percolation is an extension of a percolation model which has interesting applications to the resilience of networks under random damage.
In this model, the notion of node robustness is local, instead of global as in uniform $k$-core percolation.
One of the advantages of $k$-core percolation models is the validity of an analytical mathematical framework for a large class of network topologies.
We study ternary mixtures of node types in random networks and show the presence of a new type of critical phenomenon.
This scenario may have useful applications in the stability of large scale infrastructures and the description of glass-forming systems.
}

\section{Introduction}
\label{sec:1}
Percolation, with its many modifications and extensions, is a problem with a venerable past and many applications in the most diverse disciplines \cite{stauffer1994}.
With the rapid development of network science, several new problems and models have been introduced.
Among the different directions, {\kcore} percolation constitutes a development which, in spite of its somewhat simple definition, is able to encapsulate a number of interesting problems which can often be approached with a robust mathematical formalism.
Given a network, a {\kcore} is defined as the subnetwork where each node has at least $k$ neighbours in the same subnetwork.
A {\kcore} can equivalently be defined as the subnetwork remaining after a culling process consisting in recursively removing all the nodes with degree lower than $k$.
{\kcore} percolation has applications in many different disciplines including jamming \cite{schwarz2006}, neural networks \cite{Chatterjee2007}, granular gases \cite{alvarez2007}, evolution \cite{klimek2009}, social sciences \cite{kitsak2010} and the metal-insulator transition \cite{cao2010}.
The application of {\kcore} percolation we are interested in here, however, refers to the stability of a network under random damage.
A fraction $(1-p)$ of nodes is removed (together with the adjacent edges) and the size $M_k$ of the largest {\kcore} cluster is studied as a function of $p$.
The {\kcore} strength $M_k$ can vanish either continuously or discontinuously, a scenario that can be described in terms of phase transitions (second and first order, respectively) and can be important in the stability of large scale infrastructures \cite{cohen2000,dorogovtsev2006}.
A recent extension of {\kcore} percolation, named heterogeneous {\kcore} (HKC) percolation considers the threshold $k_i$ as a local property, and allows therefore the mixing of different critical phenomena.
An analytical formalism has been introduced to approach this model on networks and binary mixtures of thresholds $k$ have been investigated \cite{baxter2011,cellai2011,cellai2012}.
On a wide class of networks, this model displays phase diagrams which are topologically equivalent to the ones calculated for a recently introduced spin model of glass-forming systems \cite{sellitto2010,sellitto2012}.
This model can be seen as a heterogeneous development of the Fredrickson-Andersen (FA) model of facilitated spins, where the facilitation consists of a local constraint in the number of spins down in order for the considered spin to be able to flip \cite{fredrickson1984,sellitto2010}.
It has been shown that this model reproduces characteristic signatures of glass-forming systems called \emph{glass-transition singularities}, which correspond to distinctive critical phenomena for the appropriate choice of the parameters \cite{goetze2009,sellitto2012}.
One of the glass-transition singularities which has not yet been explored  is the so-called $A_4$ singularity, characterised by the coalescence of two critical points.
As the HKC model appears to reproduce all the relevant critical phenomena for binary mixtures, it is interesting to look for this singularity as well.
The purpose of this paper, therefore, is to investigate, for the first time, ternary mixtures in HKC
percolation and calculate the critical point corresponding to an $A_4$  singularity.

In Section 2 we present the analytical formalism used here (from \cite{baxter2011}), in Section 3 we sketch the behaviour of the model for binary mixtures, in Section 4 we show a case of ternary mixture with an $A_4$ singularity and in Section 5 we give the conclusions.

\section{Formalism}
\label{sec:2}
Let $k_i$ be the {\kcore} threshold of the node $i$.
A fraction $(1-p)$ of nodes are randomly removed: the problem consists in calculating the size of the HKC (if it exists).
We consider the configuration model of random networks, defined as the maximally random network with a given degree distribution $P(q)$.
The configuration model has the property of being locally tree-like, {\ie} the number of finite loops vanishes for infinite networks.
This property allows to consider, in an infinite network,  the HKC equivalent to the {\kisubtree}, defined as the tree in which, as we traverse it, each encountered vertex has at least $k_i-1$
child edges.
Then we can write a self-consistent  equation for $Z$, the probability that a randomly chosen node is the root of a {\kisubtree} \cite{baxter2011}:
\begin{eqnarray}
	Z &=& p r \sum_{q=k_a}^{\infty} \frac{qP(q)}{\langle q \rangle}
\sum_{l=k_a-1}^{q-1} {q-1 \choose l} Z^l (1-Z)^{q-1-l} +\nonumber\\
	&& + p s\sum_{q=k_b}^{\infty}  \frac{qP(q)}{ \langle q \rangle}
\sum_{l=k_b-1}^{q-1} {q-1 \choose l} Z^l (1-Z)^{q-1-l}  +\nonumber\\
	&& + p(1-r-s)\sum_{q=k_c}^{\infty} \frac{qP(q)}{ \langle q \rangle}
\sum_{l=k_c-1}^{q-1} {q-1 \choose l} Z^l (1-Z)^{q-1-l} .
	\label{eq:Z-equation}
\end{eqnarray}
where the three thresholds $\k=(k_a,k_b,k_c)$ are randomly assigned to nodes with probability $r$, $s$ and $(1-r-s)$, respectively.
In this paper we always assume $k_i\ge 2$, so we do not need to consider the case where there may be finite clusters in the HKC.
Due to the absence of finite loops, no finite HKC clusters can exist if $k_i\ge 2$ for all nodes $i$ \cite{dorogovtsev2006,baxter2011}.

We can then write the probability $M_{abc}$ that a randomly chosen node is in the HKC, for a mixture of three types of nodes:
\begin{eqnarray}
  M_{abc}(p)  &=& p r\sum_{q=k_a}^{\infty} P(q) \sum_{l=k_a}^{q} {q \choose l} Z^l(1-Z)^{q-l}+\nonumber\\
  && p s\sum_{q=k_b}^{\infty} P(q) \sum_{l=k_b}^{q} {q \choose l} Z^l(1-Z)^{q-l}+\nonumber\\
  && p (1-r-s)\sum_{q=k_c}^{\infty} P(q) \sum_{l=k_c}^{q} {q \choose l} Z^l(1-Z)^{q-l}.
  \label{eq:Mabc}
\end{eqnarray}

\section{Binary mixtures in heterogeneous $k$-core percolation}
\label{sec:3}
\begin{figure}[htb]
\sidecaption
	\includegraphics[width=\columnwidth]{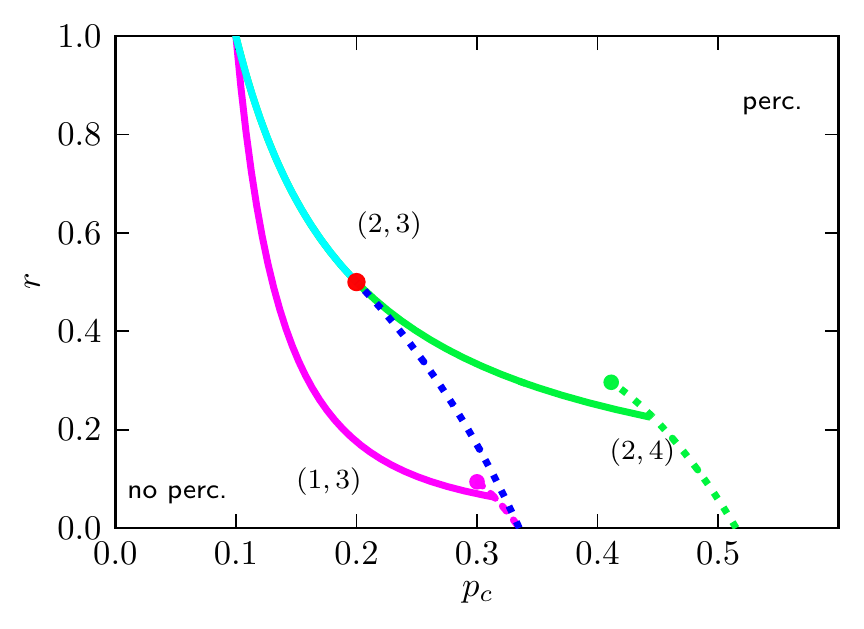}
	\caption{
		Comparison of the phase diagrams of the cases $\k=(1,3)$, $\k=(2,3)$ and $\k=(2,4)$ for \er~graphs with $z_1=10$.
		Continuous lines represent second order phase transitions, whereas dashed lines represent first order phase transitions.
		The red dot represents a tricritical point; the other dots are critical points.
		In terms of the notation given in Section 2, a binary mixture corresponds for example to $s=1-r$.}
	\label{fig:comparison-phase-diag-pr}
\end{figure}

In homogeneous {\kcore} percolation it is known that, for networks with a fast decreasing degree distribution, namely $P(q)<1/q^{\gamma}$ with $\gamma>3$ for $q\to\infty$, a randomly damaged network collapses continuously for $k\le 2$ and discontinuously for $k\ge 3$ \cite{chalupa1979,branco1993,dorogovtsev2006}.
In recent papers, it has also been shown that in the case of binary mixtures HKC percolation is characterized by a few different critical phenomena.
For example, in mixtures where the two values of $k$ are associated to continuous and discontinuous transitions, respectively, a critical point (also called $A_3$ singularity or cusp singularity) is usually observed \cite{baxter2011,cellai2012}.
That is the case, for instance, of the mixtures $\k=(1,3)$ or $\k=(2,4)$, where the line of first order transitions ends in a critical point and a line of second order transition intersects the former.
The case $\k=(2,3)$, though, is quite peculiar as the critical line exactly matches the line of first order transitions giving rise  to a tricritical point (Fig.~\ref{fig:comparison-phase-diag-pr}) \cite{cellai2011}.

\begin{figure}[htb]
\sidecaption
%\begin{center}
	\includegraphics[width=\columnwidth]{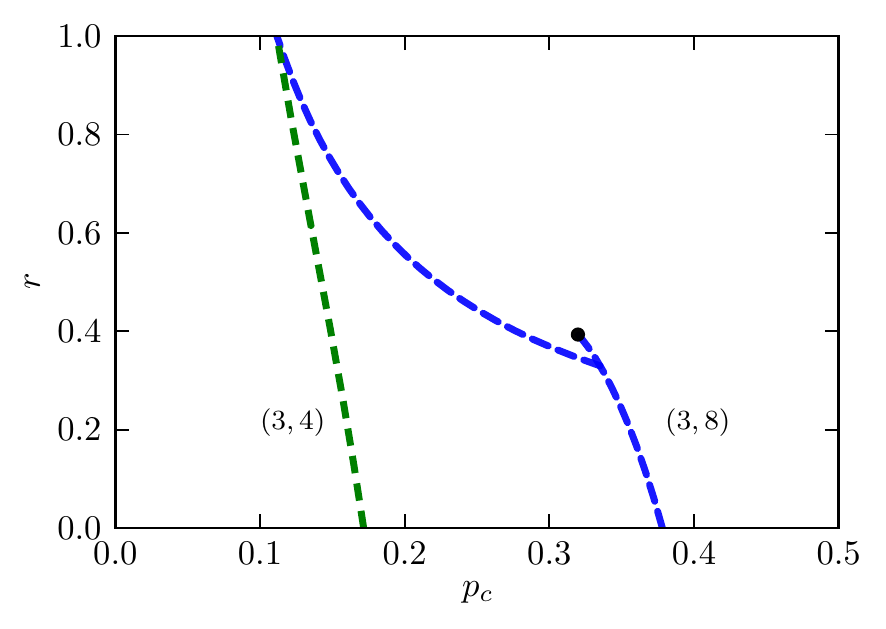}
	\caption{Phase diagram of the cases $\k=(3,8)$  and $\k=(3,4)$ for the \er~graph ($z_1=30$). 
	Symbols are as in Fig.~\ref{fig:comparison-phase-diag-pr}.
	While for $\k=(3,4)$ there is only one line, in the case $\k=(3,8)$ there are two lines of discontinuous transitions and a critical point.}
	\label{fig:er-3-8-ph-diag}
%\end{center}
\end{figure}

When both values of $k$ are characterised by a first order transition, instead, the phase diagram displays a critical point only when the two values of $k$ are different enough \cite{cellai2012}.
Fig.~\ref{fig:er-3-8-ph-diag}, for instance, shows that the mixture $\k=(3,8)$ has a critical point whereas $\k=(3,4)$ does not have one.

\section{The $A_4$ singularity}
\label{sec:4}
From the analogy between HKC percolation and facilitated spin models, it is interesting to investigate whether the singularities observed in models of the glass-forming systems are also present in HKC percolation.
In particular, the fact that all the critical phenomena observed so far in HKC percolation are in the same class of universality as the ones observed in the FA model with heterogeneous facilitation is  quite remarkable \cite{sellitto2012,cellai2012}.
The MCT of glass-forming systems also predicts an ``$A_4$ singularity'', meaning the coalescence of a critical point into a line of first order phase transitions \cite{goetze2009}.
The $A_4$ singularity is also named  a swallow-tail bifurcation.
As this scenario is associated with a three-parameter theory, it is reasonable to expect that the corresponding singularity in our HKC model can be observed in a three-component mixture.
Moreover, we have seen in Section 3 that binary mixtures with $k_i\ge 3$ are either characterised by a single line of first order transitions or two lines of first order transitions with a single critical point.
It is therefore natural to consider a ternary mixture which interpolates between the two regimes.
We now show that an $A_4$ singularity is indeed present in the ternary mixture $\k=(3,5,8)$ on {\er} graphs.

%The critical phenomena observed in the heterogeneous FA model correspond in turn to singularities in the mode coupling theory (MCT) of glass-forming systems \cite{arenzon2012,goetze2009}.
%One of the higher-order glass singularities observed in MCT is defined as an $A_4$ singularity according to Arnold \cite{arnold}.
%It consists of the intersection point between two distinct lines of first order transition, where two $A_3$ singularities coalesce \cite{goetze,goezte2002}.
%It is therefore meaningful to look for an $A_4$ singularity in HKC percolation as well.
%
%In MCT, the $A_4$ singularity is observed in the case $F_{123}$, which is characterised by three parameters, and it is defined by the vanishing of the third derivative of the memory kernel \cite{goetze}. 

\begin{figure}[htb]
\sidecaption
%\begin{center}
	\includegraphics[width=\columnwidth]{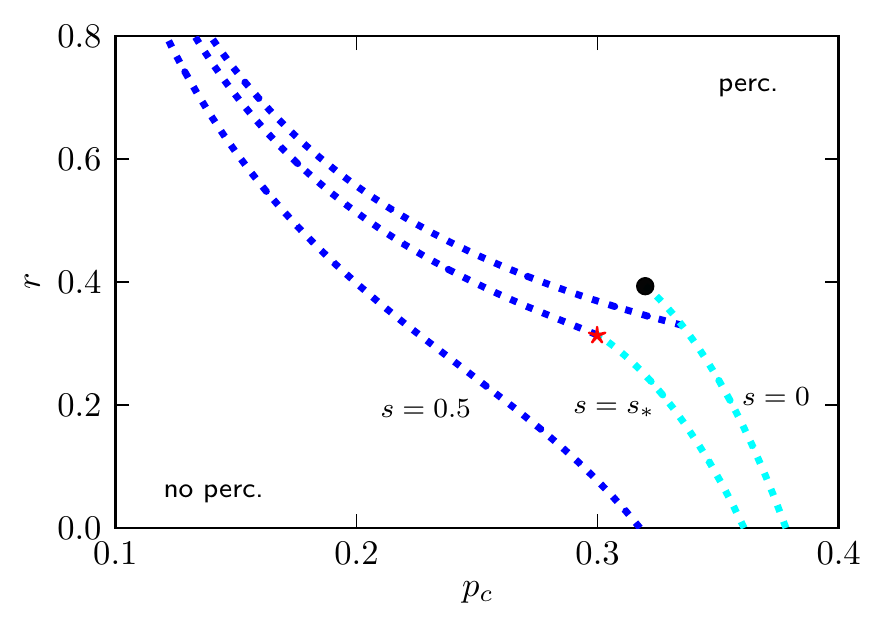}
	\caption{Phase diagram of the mixture $\k=(3,5,8)$ for \er~graphs ($z_1=30$).
	$r$ and $s$ are the fractions of nodes of type $3$ and $5$, respectively, and $p_c$ is the fraction of undamaged nodes at a phase transition.
	Lines of discontinuous transitions are plotted in the plane $(p,r)$ at three fixed values of $s$.
	In the absence of type 5 ($s=0$), there are two lines of first order transitions.
	One of them ends in a critical point (black dot).
	At the critical fraction $s_*$, the two lines touch at a $A_4$ singularity (red star).
	For $s>s_*$, there is only a single line of discontinuous transitions.}
	\label{fig:er-3-5-8-ph-diag}
%\end{center}
\end{figure}

Equation~(\ref{eq:Z-equation}) can be re-written as 
\begin{equation}
	f_{358}(Z)=\frac{1}{p},
\end{equation}
where
\begin{equation}
	f_{358}(Z)=\frac{1}{Z}\left\{  1-e^{-z_1Z}\left[ 1+r z_1Z +s\sum_{n=1}^3\frac{(z_1Z)^n}{n!} + (1-r-s)\sum_{n=1}^6\frac{(z_1Z)^n}{n!}  \right]  \right\}.
\end{equation}
At every fixed fraction of remaining nodes $p$, depending on the values of the parameters $r$ and $s$, $f_{358}$ has one or two maxima in $Z$.
If the second maximum is higher than the first one, there is only one first order transition between a percolating and a non-percolating HKC.
If the first maximum is higher than the second one, there are two first order transitions: one between a high-$k$ and a low-$k$ phase and another towards a collapsing HKC.
Typically, the second maximum disappears at a critical point, which can be found by imposing the condition:
\begin{equation}
	f'_{358}(Z) = f''_{358}(Z) = 0.
\end{equation}
This defines a locus in the  $(r,s)$ plane corresponding to a line of critical points.
At high values of $s$, however, the critical point disappears and only a single line of first order transitions survives.
This is due to the fact that the 5-nodes interpolate between the other two values of $k$ and there is no  critical point  in either  the mixture $\k=(3,5)$ or $\k=(5,8)$.
The condition of an $A_4$ singularity corresponds to the onset of this behaviour and is defined by the condition:
\begin{equation}
	f'_{358}(Z) = f''_{358}(Z) = f'''_{358}(Z) = 0,
\end{equation}
which yields
\begin{equation}
	r_* = 0.3140687639806 \qquad s_* = 0.1831697392197
\end{equation}
which is the position of the $A_4$ singularity on the phase diagram.
Fig.~\ref{fig:er-3-5-8-ph-diag} shows the phase diagram in the plane $(p,r)$ for a few values of $s$.

The critical exponent $\beta$ is defined by the vanishing of the order parameter in approaching a phase transition: $M_{358}(p)-M_{358}(p_c) \sim (p-p_c)^{\beta}$.
As in binary mixtures, the critical exponent at the discontinuous phase transition is $\beta =1/2$, whereas it becomes  $\beta = 1/3$ at the critical point.
In the vicinity of the $A_4$ singularity, we have $(p-p_*) \sim f_{358}(Z_*)-f_{358}(Z) \sim (Z-Z_*)^4$, from which it follows that $M_{358}(p)-M_{358}(p_*) \sim (p-p_*)^{1/4}$, as $M_{358}$ is linear in $p$ and $Z$ in the vicinity of the transition.
Thus, the values of the critical exponent $\beta$ can be summarized as follows:
\begin{equation}
	\beta = \left\{%
	\begin{array}{lll}
		1/2 & \textrm{hybrid transition} & r\neq r_c,  s\neq s_*\\
		1/3 & \textrm{critical point} & r=r_c, s\neq s_*\\
		1/4 & \textrm{$A_4$ singularity} & r=r_*, s= s_*\\
	\end{array}
	\right.
	\label{eq:3-5-8-beta-dxt}
\end{equation}
The unique value of $\beta=1/4$ identifies the critical point as a $A_4$ singularity.
This phenomenon has an interesting interpretation.
In a  networks formed by types of nodes of very different fragilities (as in $k_a=3$, $k_c=8$), the collapse due to random damage can be catastrophic, with the presence of multiple discontinuous transitions.
The introduction of nodes of intermediate fragility ($k_b=5$) can weaken the critical point up to disappearance, making a significant region of the phase diagram stable (Fig.~\ref{fig:er-3-5-8-ph-diag}).
This purely mathematical remark could be useful in principle in planning the stability of large scale infrastructures.
Another interesting application is in modelling the glass transition.
The presence of $A_4$ singularities is predicted by the mode-coupling theory of glass-forming systems \cite{goetze2009}.
Due to the strong similarities between HKC percolation and facilitated spin models, it is reasonable to expect that this singularity should also be present in suitable ternary mixtures of facilitation parameters in such spin models \cite{sellitto2010,sellitto2012}.

\section{Conclusions}
\label{sec:5}
In this paper we have applied the formalism of heterogeneous {\kcore} percolation to ternary mixtures of thresholds $k$.
In particular, we have calculated the phase diagram of a ternary mixture for {\er} graphs.
This mixture displays a characteristic $A_4$ singularity, {\ie} a critical phenomenon characterised by the merging of a critical point with a distinct line of discontinuous transitions.
A peculiarity of this point is the change in the critical exponent $\beta$ of the order parameter, which uniquely assumes the value $1/4$.
The characteristics of the studied phase diagrams may give useful information in designing  large scale infrastructures to be resilient to random damage.
This singularity is also predicted by models of the glass transition \cite{sellitto2012} and our model appears to be in the same universality class of kinetic spin models with heterogeneous facilitation.

\begin{acknowledgement}
We acknowledge useful discussions with Mauro Sellitto.
This work has been funded by Science Foundation Ireland, grants: 11/PI/1026 and 06/MI/005.
\end{acknowledgement}

\end{document}